\documentstyle[11pt,draft]{article}
\def\flecha{\mathop{\longrightarrow}}

\author{A. Mondrag\'on and E. Hern\'andez\thanks{This work was partially 
supported by CONACYT (M\'exico) under contract No. 4964-E9406}\\
Instituto de F\'{\i}sica, UNAM\\
Apartado Postal 20-364, 01000 M\'exico D.F. MEXICO}
\title{Berry Phase of a Resonant State
}
\date{ 
}

\begin{document}

\maketitle
\begin{abstract}
We derive closed analytical expressions for the complex Berry phase of an
open quantum system in a state which is a superposition of resonant states
and evolves irreversibly due to the spontaneous decay of the metastable
states. The codimension of an accidental degeneracy of resonances and the
geometry of the energy hypersurfaces close to a crossing of resonances
differ significantly from those of bound states. We discuss some of the
consequences of these differences for the geometric phase factors, such as:
Instead of a diabolical point singularity there is a continuous closed line
of singularities formally equivalent to a continuous distribution of 
`magnetic' charge on a diabolical circle; different classes of topologically 
inequivalent non-trivial closed paths in parameter space, the topological 
invariant associated to the sum of the geometric phases, dilations of the 
wave function due to the imaginary part of the Berry phase and others.
\end{abstract}


\section{Introduction.}

During the last ten years the geometric phase factors arising in the
adiabatic evolution of quantum systems\cite{one} have been the subject of
many investigations\cite{two}. Most of the early literature was concerned
with the geometric phase factors of closed systems driven by Hermitian
Hamiltonians\cite{three}. More recently there has been substantial interest
in the complex geometric phase acquired by the eigenstates of open quantum
systems. This problem arises naturally in connection with various
experiments which, by their very essence, require the observation of the
Berry phase in metastable states ( also called resonant or Gamow states).
Dattoli et al.\cite{four} studied the Berry phase in the optical supermode
propagation in a free electron laser, which is a classical system described
by a Schr\"odinger-like equation with a non-Hermitian Hamiltonian. The
measurement of the geometric phase in atomic systems with two energy levels,
one of which at least is metastable, was also described in terms of a
non-Hermitian Hamiltonian by Miniatura et al.\cite{five}. The validity of
the adiabatic approximation for dissipative, two level systems driven by
non-Hermitian Hamiltonians was examined by Nenciu and Rasche\cite{six}, and
by Kvitsinsky and Putterman\cite{seven} who also established that, in this
case, the Berry phase is complex. A higher order adiabatic approximation for
two level non-Hermitian Hamiltonians was proposed by C.P. Sun\cite{eight},
who also showed that the holonomy structure associated to the Berry phase
factor of the non-Hermitian case is the non-unitary generalization of the
holonomy structure of the Hermitian case. In a previous paper, we gave
closed analytical expressions for the geometric phase of true resonant states
\cite{nine} defined as energy eigenstates of a Hermitian Hamiltonian which
satisfy purely outgoing wave boundary conditions at infinity\cite{ten}, and
pointed out some of the mathematically interesting and physically relevant
properties resulting from the non-selfadjointness of the problem. In this
connection, later, we showed that the codimension of the accidental
degeneracy of $n$ resonances and the topological structure of the energy
surfaces close to a crossing of resonances differ significantly from those
of bound states\cite{eleven}. By means of a numerical analysis of the
experimental data on the $2^{+}$ doublet of resonances with $T=0,1$ in the
energy spectrum of the nucleus $^8Be$ at about $16.6$, and $17MeV$, we
showed in a realistic example, that a true crossing of resonances mixed by
a Hermitian interaction may be brought about by the variation of only two
real independent parameters\cite{twelve}. H-H. Lauber, P. Weidenhammer and
D. Dubbers\cite{thirteen} measured the geometric phase associated to a
triple degeneracy of resonances in a flat, non-symmetric, triangular
microwave resonator.

In this paper we explore further the properties of the Berry phase of
resonant states of a Hermitian Hamiltonian with non-selfadjoint boundary
conditions. The paper is organized as follows: In section 2, we consider the
time evolution of a quantum system in a state which is a superposition of
bound and unbound states evolving in the presence of an external field of
force which changes slowly with time. By means of an expansion of the wave
function in terms of bound and resonant states of the unperturbed system, we
associate a complex non-Hermitian matrix to the time evolution of the bound
and resonant states evolving under the action of the time dependent
perturbation. Accidental degeneracy of two resonances and the properties of
the energy hypersurfaces at the crossing of resonances are discussed in
section 3. In section 4 we derive general expressions for the Berry phase of
resonant states and discuss some of its properties. In section 5 we briefly
discuss the validity of the adiabatic approximation. The paper ends in
section 6 with a short summary of results and some conclusions.

\section{Adiabatic Mixing of Resonant States}

Let us consider the adiabatic time evolution of a quantum system in a state
which is a superposition of unstable states moving in some strong external
field of force which changes slowly with time. In order to have some
concrete example in mind, although a very hypothetical one, we may think of
an $^8Be$ nucleus which has only unstable energy eigenstates moving in the
field of a heavy doubly magic nucleus, like $^{208}Pb,$ in a peripheral
collision in which the distance between the two nuclei is never much smaller
than the sum of the nuclear radii. In a semiclassical treatment of the
collision, when the centers of the nuclei move along given classical
trajectories, the parameters in the nucleus-nucleus interaction change with
time\cite{fifteen}. Another example could be a highly excited Hydrogen atom
in strong external, crossed electric and magnetic fields. This system has no
bound states, it has only resonances\cite{sixteen}; when the external fields
change slowly with time, the parameters in the atom-field interaction change
with time.

The evolution of the unstable quantum system under the influence of the
external perturbation is governed by the time dependent Schr\"odinger
equation

\begin{equation}
\label{uno}i\hbar \frac{\partial \Psi }{\partial t}=H\Psi . 
\end{equation}
The Hamiltonian $H$ is the sum of the time-independent Hamiltonian $H_0$
describing the evolution of the unperturbed unstable system plus a
perturbation term $H_1$ parametrized in terms of some set of external
parameters$\left\{ C_1,C_2....C_N\right\} $.

The energy eigenfunctions of the unperturbed Hamiltomian are the solutions
of the equation

\begin{equation}
\label{dos}H_0\varphi _m(\xi _i)={\cal E}_m\varphi _m(\xi _i) 
\end{equation}
plus the appropriate boundary conditions.

Assuming that in the absence of perturbation the unstable nucleus decays
spontaneously in two stable nuclear clusters, the unperturbed energy
eigenfunctions may be written as cluster model wave functions\cite{seventeen},

\begin{equation}
\label{tres}\varphi _m(\xi _i,\xi _j)={\cal A}\left\{ \phi _A(\xi _i)\phi
_B(\xi _j)\frac{u_{AB}(r_{AB})}{r_{AB}}{\cal Y}_J^M(\hat r_{AB})\right\}, 
\end{equation}
where $\phi _A$ and $\phi _B$ are the wave functions of the clusters A and
B, $u_{AB,m}(r_{AB})$ is the radial part of the wave function of the
relative motion of the two clusters ${\cal Y}_J^M(\hat r_{AB})$ is a
spherical harmonic, and ${\cal A}$ is the antisymmetrizer. In our example, 
$\varphi _m$ would be the eigenfuction of a state of $^8Be$ which decays
spontaneously in two $^4He$ clusters. In this case, $u_{AB,m}(r_{AB})$ is a
Gamow function which vanishes at the origen and behaves as a purely outgoing
wave for large values of the relative distance $r_{AB,}$

\begin{equation}
\label{cuatro}\frac 1{u_{AB,m}(r_{AB})}\left( \frac{du_{AB,m}(r_{AB})}{dr}
\right) \flecha_{r_{AB}\rightarrow \infty }ik_m. 
\end{equation}
$k_m$ is the complex wave number. This boundary condition is not
self-adjoint, in consequence the Gamow states are not orthogonal in the
usual sense, and the energy eigenvalues are complex, with $Re{\cal E}_m>0$
and $Im{\cal E}_m< 0.$ Nevertheless, bound and resonant states form a
biorthonormal set with their adjoints, which may be extended by a continuum
of suitably chosen scattering states of complex wave number $k$ to form a
complete set in which any square integrable function may be expanded. The
rules of normalization, orthogonality and completeness satisfied by the
eigenstates of the Schr\"odinger equation with local and non-local
potentials belonging to complex eigenvalues with $ImE_n<0,$ (Gamow or
resonant states) are given in A. Mondrag\'on, E.\ Hern\'andez and J.M.
Vel\'azquez-Arcos\cite{ten}. A clear exposition of the properties of
resonant states may be found in the textbook by A. B\"ohm\cite{fourteen}.

Since we are interested in the time evolution of the system in a state which
is a superposition of unstable states, we make an expansion of the wave
function $\Psi $ in terms of bound and resonant states of $H_0,$

\begin{equation}
\label{cinco}\Psi =\sum_ma_m(t)\varphi _m(\xi _i)+\int_cb(k;t)\varphi
^{(+)}(k;\xi )dk 
\end{equation}

In general, the index $m$ runs over bound and resonant states. The
scattering states $\varphi ^{(+)}(k;\xi)$ of complex wave number $k$ 
are defined by analytic continuation\cite{ten}. The integration contour C 
in the complex wave number plane is a straight line with slope -1, that 
goes through the origin. In this way we separate out of the continuum the
resonance contribution to $\Psi$, which should be dominant for energies
close to the real part of the complex energy eigenvalues.

Substitution of (\ref{cinco}) in (\ref{uno}) gives the set of coupled
equations

\begin{equation}
\label{seis} 
\begin{array}{c}
\frac{da_m(t)}{dt}=-\frac i\hbar {\cal E}_ma_m(t)-\frac i\hbar
\sum_n<\varphi _m|H_1|\varphi _n>a_n(t) \\ -\frac i\hbar \int_c<\varphi
_m|H_1|\varphi ^{(+)}(k)>b(k;t)dk 
\end{array}
\end{equation}
and a similar expression for $\frac{db(k,t)}{dt}.$

When the interactions are time reversal invariant, the dual of the complex
Gamow function $u_{AB,m}(r_{AB})$ is the same function\cite{ten}. Since the
bound state cluster wave functions $\phi _A$ and $\phi _B$ are real, the
dual of the resonant state wave function $\varphi _m(\xi _i)$ is the same
complex function $\varphi _m(\xi _i).$ When the interactions are not time
reversal invariant, the Gamow function and its dual are not the same
function. We will use the notation $|\varphi _m(\xi _i)>$for the Gamow
function, and $<\varphi _m(\xi _i)|$ for its dual. Hence, the matrix element
of the perturbation term $H_1$ taken between bound or resonant states of the
unperturbed system is given by

\begin{equation}
\label{siete}\left\langle \varphi _m\left| H_1\right| \varphi
_n\right\rangle =\int \int \langle \varphi _m(\xi _i)|H_1|\varphi _n(\xi
_i)\rangle d\xi _1\cdot \cdot \cdot \cdot d\xi _i 
\end{equation}
The matrix ${\bf H}$ with matrix elements

\begin{equation}
\label{ocho}{\bf H}_{m,n}={\cal E}_m\delta _{m,n}+\left\langle \varphi
_m\left| H_1\right| \varphi _n\right\rangle 
\end{equation}
is, in general, non-symmetric, complex and non-Hermitian. When the forces
acting on the system are time reversal invariant, ${\bf H}$ is symmetric,
but non-Hermitian. The dependence of $H_1$ on the external parameters is
inherited by ${\bf H}$ which is also parametrized in terms of the same set 
$\left\{ C_1,C_2,....C_n\right\} $ of external parameters.

In the following we will be concerned with the adiabatic evolution of a
system with two resonant states very close in energy which are strongly
mixed by the Hermitian interaction $H_1.$ In our hypothetical example, these
could be the two $2^{+}$ states of $^8Be$ with $T=0,1$ at $E_1=16.622MeV$
and $E_2=17.01MeV$. The contribution of the non-resonant background integral
over the continuum of scattering functions will play no part in the
following discussion. Therefore, to ease the notation, we will disregard the
contribution of the background scattering functions. In this approximation,

\begin{equation}
\label{nueve}H\Psi =\sum_m\left[ \sum_n{\bf H}_{m,n}a_n(t)\right] |\varphi
_m(\xi )>. 
\end{equation}

Then, the set of eqs.(\ref{seis}) reduces to

\begin{equation}
\label{diez}\dot a_m(t)=-\frac i\hbar \sum_n{\bf H}_{m,n}a_n(t) 
\end{equation}
where $\dot a_m(t)$ is the time derivative of $a_n(t).$

The complex energy eigenvalues associated to the resonant states in the
presence of the perturbation will arise as the eigenvalues of the
non-Hermitian matrix ${\bf H}$ associated, through eq.(\ref{diez}), with the
decaying process.

\section{Accidental Degeneracy of Resonances}

Non-trivial geometric phase factors of the energy eigenvectors or
eigenfunctions are related to the occurrence of accidental degeneracies of
the corresponding eigenvalues\cite{one,nineteen}. In the previous
section we associated a complex $n\times n$, non-Hermitian matrix ${\bf H}$
to an open quantum system with $n$ resonances and resonant states. In the
absence of symmetry, degeneracies are called accidental for lack of an
obvious reason to explain why two energy eigenvalues, ${\bf E}_1$ and
${\bf E}_2$, of $\bf H$ should coincide. However, if the matrix $\bf H$
is embedded in a population of complex non-Hermitian matrices
$\left\{ {\bf H} (C_1,C_2,.....,C_N)\right\}$ smoothly parametrized by $N$ 
external parameters $(C_1,C_2,....,C_N)$, degeneracy in the absence of
symmetry is a geometric property of the hypersurfaces representing the real or 
complex eigenvalues of $\bf H$ in a $(N+2)-$dimensional Euclidean space with
cartesian coordinates $(C_1,C_2,.....,C_N,Re{\bf E},Im{\bf E})$. In contrast
with the case of Hermitian matrices, complex non-Hermitian matrices with
repeated eigenvalues can not always be brought to diagonal form by a
similarity transformation. This feature leads to a richer variety of
possibilities. Thus, it will be convenient to examine first the accidental
degeneracy of two resonances and the topology of the energy surfaces close to
a crossing of resonances in parameter space.

\subsection{Degeneracy of two resonances.}

Since we are interested in the possibility of an accidental degeneracy in a
system with two resonant states strongly mixed by a Hermitian interaction,
all other bound or resonant eigenstates being non-degenerate, we may suppose
that we already know the correct eigenvectors of ${\bf H}$ for all the real
and complex eigenenergies ${\bf E}_s$, except for the two the crossing of
which we want to investigate. Using for this two states two vectors which
are not eigenvectors but which are orthogonal to each other and to all other
eigenvectors, we obtain a complete basis to represent ${\bf H}$. In this
basis ${\bf H}$ will be diagonal except for the elements ${\bf H}_{12}$ and
${\bf H}_{21}$. The diagonal elements ${\bf H}_{11}$ and ${\bf H}_{22}$ will,
in general, be non-vanishing and different from each other. There is no loss
of generality in this supposition, since any complex matrix ${\bf H}$ may be
brought to a Jordan canonical form by means of a similarity transformation%
\cite{eighteen}. When the eigenvalues are equal, ${\bf H}%
_{2\times 2}$ is either diagonal or equivalent to a Jordan block of rank
two. Hence, in the following, we need consider only the conditions for
degeneracy of the submatrix ${\bf H}_{2\times 2}$.

It will be convenient to measure the resonance energies, ${\bf E}_1$and $%
{\bf E}_2,$ from the centroid, ${\cal E}$, of the diagonal terms in ${\bf H}
_{2\times 2}$, then,

\begin{equation}
\label{once}{\bf H}_{2\times 2}={\cal E}{\bf 1}+{\cal H,} 
\end{equation}
the traceless $2\times 2$ matrix ${\cal H}$ may be written in terms of the
Pauli-matrix valued vector $\vec \sigma =(\sigma _1,\sigma _2,\sigma _3)$ as

\begin{equation}
\label{doce}{\cal H}=\left( \vec R-i\frac 12\vec \Gamma \right) \cdot \vec
\sigma, 
\end{equation}
where $\vec R$ and $\vec \Gamma $ are real vectors with cartesian components 
$\left( X_1,X_2,X_3\right) $and $\left( \Gamma _1,\Gamma _2,\Gamma _3\right) 
$given by

\begin{equation}
\label{trece}X_i-i\frac 12\Gamma _i=\frac 12\left[ Tr.\left( {\cal H}\sigma
_i\right) \right] , 
\end{equation}
when the forces acting on the system are time reversal invariant, $X_2$ and
$\Gamma _{2}$vanish.

In the absence of more specific information about the external parameters
$C_i$, we will parametrize ${\cal H}$ in terms of $\vec R$ and $\vec \Gamma$
, according to (\ref{once}) and (\ref{doce}). It will be asumed that 
$X_i=X_i(C_1,C_2,....C_N)$ are bijective and define a homeomorphism of the
manifold of the external parameters onto the manifold of the complex,
non-Hermitian $2\times 2$ matrices ${\bf H}_{2\times 2}.$

From (\ref{doce}), the eigenvalues of ${\cal H}$ are given by

\begin{equation}
\label{catorce}\epsilon =\mp \sqrt{\left( \vec R-i\frac 12\vec \Gamma
\right) ^2} 
\end{equation}
and the corresponding eigenvalues of ${\bf H}$ are

\begin{equation}
\label{quince}{\bf E}_{1,2}={\cal E}\mp \epsilon . 
\end{equation}

Then, ${\bf E}_1$ and ${\bf E}_2$ coincide when $\epsilon $ vanishes. Hence,
the condition for accidental degeneracy of the two interfering resonances
may be written as

\begin{equation}
\label{dieciseis}\sqrt{\left( \vec R_d-i\frac 12\vec \Gamma _d\right) ^2}=0 
\end{equation}

Since real and imaginary parts should vanish, we get the pair of equations

\begin{equation}
\label{diecisiete}R_d^2-\frac 14\Gamma _d^2=0, 
\end{equation}

\noindent
and

\begin{equation}
\label{dieciocho}\vec R_d\cdot \vec \Gamma _d=0. 
\end{equation}

These equations admit two kinds of solutions corresponding to  ${\bf H}
_{2\times 2}$ being or not being diagonal at the degeneracy:

i) When both $\vec R_d$ and
$\vec \Gamma _{d}$ vanish, eqs. (\ref{diecisiete}) and (\ref{dieciocho})
define a point in parameter space, ${\bf H}$ is diagonal at the degeneracy
and the submatrix ${\bf H}_{2\times 2}$ has two cycles of eigenvectors of
lenght one. The two complex eigenvalues of ${\bf H}$ which become degenerate
migrate to the real axis where they fuse into one real positive energy
eigenvalue embedded in the continuum\cite{twentyone}. Since in this case all 
the cartesian components of $\vec R_d$ and $\vec \Gamma _d$ should vanish, the
minimum number of external parameters that should be varied to produce
a degeneracy of two resonances to form a bound state embedded in the
continuum is four or six depending on the quantum system being or not being
time reversal invariant.

ii) In the second case, when the degeneracy conditions (\ref{diecisiete})
and (\ref{dieciocho}) are
satisfied for non-vanishing $\vec R_d$ and $\vec \Gamma _d$, these equations
define a circle in parameter space. In this case ${\cal H}_d$ does not
vanish at the degeneracy and the corresponding ${\bf H}_d$ is non-diagonal
at the degeneray, its first Jordan block ${\bf H}_{2\times 2}$ is of rank
two and has one cycle of generalized eigenvectors of lenght two, all other
Jordan blocks have one cycle of lenght one. In this case the two complex
eigenvalues of ${\bf H}$ which become degenerate fuse into one repeated
complex eigenvalue ${\cal E}$. Since, the two linearly independent
conditions (\ref{diecisiete}) and (\ref{dieciocho}) should be satisfied
for non-vanishing values of ${\vec R}_d$ and ${\vec \Gamma}_d$, at least
two real, linearly independent parameters should be varied to produce a
rank two degeneracy of resonances. Hence, the codimension of a second rank
degeneracy of resonances is two, independently of the time reversal invariance
character of the interaction.

If we had considered the possibility of ${\bf H}$ having more than two equal
eigenvalues, we could have had degeneracies of higher rank\cite
{eleven,eighteen}. Triple degeneracies of resonances in the absence of
symmetry have already been observed by Dubbers\cite{thirteen} in a metallic
cavity excited by a wave guide. Explicit expressions for the codimension of
a resonance degeneracy of $\upsilon $ resonance eigenenergies for 
${\upsilon \geq 2}$, have been given by Hern\'andez and 
Mondrag\'on\cite{eleven}.

\subsection{Energy surfaces in parameter space.}

Close to a degeneracy, the energy difference, $\epsilon $, is given by eq.
(\ref{catorce}). From this expression, the real and imaginary parts of 
$\epsilon $ are

\begin{equation}
\label{diecinueve}Re\epsilon _{\pm }=\pm \left[ \frac 12\left\{ \left[
\left( R^2-\frac 14\Gamma ^2\right) ^2+\left( \vec R\cdot \vec \Gamma
\right) ^2\right] ^{\frac 12}+\left( R^2-\frac 14\Gamma ^2\right) \right\}
\right] ^{\frac 12}, 
\end{equation}
and

\begin{equation}
\label{veinte}Im\epsilon _{\pm }=\mp \left[ \frac 12\left\{ \left[ \left(
R^2-\frac 14\Gamma ^2\right) ^2+\left( \vec R\cdot \vec \Gamma \right)
^2\right] ^{\frac 12}-\left( R^2-\frac 14\Gamma ^2\right) \right\} \right]
^{\frac 12}. 
\end{equation}

These equations define two hypersurfaces in parameter space. We are
interested in the shape of the energy surfaces in the neighbourhood of a
crossing resulting from the accidental degeneracy of two resonant states.

First, we shall consider the accidental degeneracy of two resonances leading
to a rank one degenerate Hamiltonian matrix. In this case, the conditions
for accidental degeneracy, eqs.(\ref{diecisiete}) and (\ref{dieciocho}), are
only satisfied for vanishing $\vec R$ and $\vec \Gamma .$ As shown in
section 3.1, the submatrix ${\bf H}_{2\times 2}$ should have at least four
free real linearly independent parameters to bring about a degeneracy of
this type. There are only three independent parameters in $\vec R$. Hence, 
$\vec \Gamma $ cannot be a fixed vector. Since $\vec R$ and $\vec \Gamma $
should vary independently of each other, it will be convenient to understand
eqs.(\ref{diecinueve}) and (\ref{veinte}) as defining the energy
hypersurfaces in an eight dimensional Euclidean space, ${\cal E}_8{\cal ,}$
with Cartesian coordinates $\left\{ X,Y,Z,u,v,w,Re\epsilon ,Im\epsilon
\right\} $. The coordinates $\left( X,Y,Z\right) $ are the Cartesian
components of $\vec R$, while $\left( u,v,w\right) $ are those of $\vec
\Gamma $. In this representation the crossing takes place at the origin of
coordinates, that is, where $\vec R=0$ and $\vec \Gamma =0.$

When the Hermitian and anti-Hermitian parts of ${\cal H}$ commute the
problem simplifies. From

\begin{equation}
\label{veintiuno}\left[ \vec R\cdot \vec \sigma ,\vec \Gamma \cdot \vec
\sigma \right] =i\left( \vec R\times \vec \Gamma \right) \cdot \vec \sigma , 
\end{equation}
it follows that, when $\vec R\cdot \vec \Gamma $ and $\vec \Gamma \cdot \vec
\sigma $ commute, $\vec R\times \vec \Gamma $ vanishes and $\vec R\cdot \vec
\Gamma $ is equal to $R\Gamma $. Then, the equations of the energy, (\ref
{diecinueve}) and (\ref{veinte}), take the simple form

\begin{equation}
\label{veintidos}Re\epsilon _{\pm }=\pm \left| R\right| =\pm \sqrt{%
X^2+Y^2+Z^2} 
\end{equation}
and

\begin{equation}
\label{veintitres}Im\epsilon _{\pm }=\mp \frac 12\left| \Gamma \right| =\mp
\frac 12\sqrt{u^2+v^2+w^2} 
\end{equation}

In this particularly simple case the two hypersurfaces which represent the
real and imaginary parts of the energy are double cones lying in orthogonal
subspaces with the apices at the origin.

In general $\vec R\cdot \vec \sigma $ and $\vec \Gamma \cdot \vec \sigma $
do not commute and $\vec R\times \vec \Gamma $ may be written as $R\Gamma \sin
\theta $ with sin$\theta \neq 0$. To study the behaviour of the energy
hypersurfaces close to the crossing point, we will approach the origin of
coordinates keeping $\vec R\cdot \vec \Gamma /R\Gamma =\cos \theta $ fixed
and let $R$ and $\Gamma $ go to zero in such a way that the ratio $R/\Gamma $
is constant. Expanding the right hand sides of (\ref{diecinueve}) and (\ref
{veinte}) in powers of $\left( \frac{2R}\Gamma \right) ^2$ when $R<\frac
12\Gamma $, or in powers of $\left( \frac \Gamma {2R}\right) ^2$ when $%
R>\frac 12\Gamma $, and keeping only the terms of lowest order, we obtain

\begin{equation}
\label{veinticuatro}Re\epsilon _{\pm }\simeq \pm \alpha \left| R\right| =\pm
\alpha \sqrt{X^2+Y^2+Z^2} 
\end{equation}

\begin{equation}
\label{veinticinco}Im\epsilon _{\pm }\simeq \mp \beta \frac 12\left| \Gamma
\right| =\mp \beta \frac 12\sqrt{u^2+v^2+w^2}
\end{equation}
where the factors $\alpha ,\beta $ are $\hat R\cdot \hat \Gamma $ and $1,$
respectively, if $R<\frac 12\Gamma ;$ when $R=\frac 12\Gamma $, $\alpha $
and $\beta $ are both equal to $\sqrt{\frac 12\hat R.\hat \Gamma };$
finally, when $R>\frac 12\Gamma ,$ $\alpha $ and $\beta $ are $1$ and $\hat
R\cdot \hat \Gamma $ respectively.

Therefore, close to a crossing of rank one, the hypersurfaces representing
the real and imaginary parts of the energy are two double cones lying in
orthogonal subspaces with their apices at the same point, which, for this
reason may be called a double diabolical point.

Now, let us consider the shape of the energy hypersurfaces close to a 
degeneracy of two resonances of rank two. In this case, the 
conditions for accidental degeneracy, eqs.(\ref{diecisiete}) and 
(\ref{dieciocho}), are satisfied for non-vanishing values of $\vec R$ 
and $\vec \Gamma $ and, as shown in 3.1, we need at least two free real
parameters in ${\cal H}$ to bring about the
degeneracy. For definiteness, we will keep the anti-Hermitian part of the
Hamiltonian matrix fixed and let the parameters of the Hermitian part of 
${\bf H}_{2\times 2}$ vary. Then, $\vec \Gamma \cdot \vec \sigma $ is a
constant matrix, $\vec \Gamma $ is a fixed vector and $\vec R$ may vary. To
simplify the notation, it is convenient to choose the OZ axis aligned with 
$\vec \Gamma $. This may be acomplished by means of a similarity
transformation of ${\cal H}$ which diagonalizes $\vec \Gamma \cdot \vec
\sigma $.

In the case under consideration, equations (\ref{diecinueve}) and (\ref
{veinte}) define two hypersurfaces in a five dimensional Euclidean space, 
${\cal E}_5$, with Cartesian coordinates $\left\{ X,Y,Z,Re\epsilon
,Im\epsilon \right\} $. The hypersurfaces representing $Re\epsilon $ and 
$Im\epsilon $ are in orthogonal subspaces and touch each other only when both 
$Re\epsilon $ and $Im\epsilon $ vanish. Therefore, the set of points in
the energy hypersurfaces corresponding to a degeneracy are all in the
subspace ${\cal E}_3$ with Cartesian coordinates $\left\{ X,Y,Z,0,0\right\}$.

The second degeneracy condition, eq.(\ref{dieciocho}), requires $\vec R_d$
to be orthogonal to $\vec \Gamma$, and defines a plane $\Pi$ in ${\cal E}_3$
and a hyperplane $\hat \Pi$ in ${\cal E}_5$ with cartesian coordinates 
$\left\{X,Y,0,Re \epsilon, Im \epsilon \right\}$. The first degeneracy
condition, eq.(\ref{diecisiete}), is the equation of a circle of radius
$\frac 12\Gamma$ in the $\Pi$-plane, which we will call the diabolical circle.

In order to have an idea of the shape of the energy hypersurfaces in the
neighbourhood of the crossing, let us consider the surface resulting from
the intersection of the $\epsilon -$hypersurface and the $\hat \Pi -$
hyperplane. The equations of this surface are obtained putting $\vec R\cdot
\vec \Gamma =0$ in (\ref{diecinueve}) and (\ref{veinte}),then

\begin{equation}
\label{veintiseis}Re\hat \epsilon _{\pm }=\pm \left[ \frac 12\left\{ \left[
\left( R^2-\frac 14\Gamma ^2\right) ^2\right] ^{\frac 12}+\left( R^2-\frac
14\Gamma ^2\right) \right\} \right] ^{\frac 12} 
\end{equation}
and

\begin{equation}
\label{veintisiete}Im\hat \epsilon _{\pm }=\mp \left[ \frac 12\left\{ \left[
\left( R^2-\frac 14\Gamma ^2\right) ^2\right] ^{\frac 12}-\left( R^2-\frac
14\Gamma ^2\right) \right\} \right] ^{\frac 12} 
\end{equation}

Hence, when $R^2\geq \frac 14\Gamma ^2$,

\begin{equation}
\label{veintiocho}Re\hat \epsilon _{\pm }=\pm \left( R^2-\frac 14\Gamma
^2\right) ^{\frac 12} 
\end{equation}
and

\begin{equation}
\label{veintinueve}Im\hat \epsilon _{\pm }=0, 
\end{equation}
the energy difference is purely real. Eq.(\ref{veintiocho}) defines
a hyperbolic cone of circular cross section in parameter space,

\begin{equation}
\label{treinta}X^2+Y^2-\left( Re\hat \epsilon \right) ^2=\frac 14\Gamma
^2 
\end{equation}

As shown in Fig. 1, the diabolical circle is at the narrowest cross section
or ``waist'' of the cone where $Re\hat
\epsilon $ vanishes and the two complex energy eigenvalues are
equal. 

Similarly, when $R^2\leq \frac 14\Gamma ^2,$

\begin{equation}
\label{treintaiuno}Re\hat \epsilon _{\pm }=0 
\end{equation}
and

\begin{equation}
\label{treintaidos}Im\hat \epsilon _{\pm }=\mp \left( \frac 14\Gamma
^2-R^2\right) ^{\frac 12} 
\end{equation}
now, the energy difference $\hat \epsilon $ is purely imaginary. This is
the equation of a sphere of radius $\frac 12$ $\Gamma ,$

\begin{equation}
\label{treintaitres}X^2+Y^2+\left( Im\hat \epsilon \right) ^2=\frac
14\Gamma ^2. 
\end{equation}

The diabolical circle is at the equator of the sphere, where the energy 
difference $\hat \epsilon $ vanishes, and the two hypersurfaces touch each
other, see Fig 2.

Off degeneracy $\left( R^2\neq \frac 14\Gamma ^2\right) $, the matrix 
${\cal H}$ that mixes the two interfering resonances is

\begin{equation}
\label{treintaicuatro}{\cal H}=\left( 
\begin{array}{cc}
\eta & \xi \\ 
\xi ^{*} & -\eta 
\end{array}
\right) . 
\end{equation}
where $\xi$ and $\eta$ are short hand for $X-iY$ and $Z-i\frac 12\Gamma$
respectively. 
${\cal H}$ has two right and two left eigenvectors and may be diagonalized
by a similarity transformation

\begin{equation}
\label{treintaicinco}K^{-1}{\cal H}K=\left( 
\begin{array}{cc}
-\epsilon & 0 \\ 
0 & \epsilon 
\end{array}
\right) , 
\end{equation}
where

\begin{equation}
\label{treintaiseis}K=\frac 1{\sqrt{2\epsilon }}\left( 
\begin{array}{cc}
\sqrt{\epsilon +\eta } & \sqrt{\epsilon -\eta } \\ \sqrt{\epsilon -\eta } 
\frac{\xi ^{*}}{\left| \xi \right| } & \sqrt{\epsilon +\eta }\frac{\xi ^{*}}
{\left| \xi \right| } 
\end{array}
\right) , 
\end{equation}

When the conditions for a degeneracy $\left( R^2=\frac 14\Gamma ^2\right) $
is satisfied, $\det K$ vanishes and the matrix $K^{-1}$ no longer exists$.$
Therefore, at the degeneracy ${\cal H}_d$ cannot be diagonalized by means of
a similarity transformation. In this case, ${\cal H}$ takes the form

\begin{equation}
\label{treintaisiete}{\cal H}_d=\frac 12\Gamma \left( 
\begin{array}{cc}
-i & e^{-i\phi } \\ 
e^{i\phi } & i 
\end{array}
\right) 
\end{equation}

This matrix has only one right eigenvector $|{\cal E}_d\rangle $ and one
left eigenvector $\langle $${\cal E}_d|$, belonging to the eigenvalue 
$\epsilon _d=0$. It also has one generalized right eigenvector $|{\cal \vec E}
_d\rangle $ and one generalized left eigenvector $\langle {\cal \vec E}_d| $
belonging to the same eigenvalue $\epsilon _d=0$, and such that

\begin{equation}
\label{treintaiocho}H_d|{\cal \vec E}_d\rangle =0|{\cal \vec E}_d\rangle 
+|{\cal E}_d\rangle 
\end{equation}
and a similar expression for $\langle $${\cal \vec E}_d|$. In consequence,
the eigenvectors of ${\cal H}_d$ and ${\bf H}_d$ do not form a complete
basis. We may add the generalized eigenvector to the set of
eigenvectors to have a complete basis. Then, any vector may be expanded in
this basis. However, one should keep in mind that, although 
$|{\cal E}_d\rangle$ and $|{\cal \vec E}_d\rangle$ are orthogonal to all left
eigenvectors belonging to all the other non-degenerate eigenvalues, the
orthonormality rules for the degenerate eigenvectors are

\begin{equation}
\label{treintainueve}\langle {\cal E}_d|{\cal E}_d\rangle =\langle {\cal 
\vec E}_d|{\cal \vec E}_d\rangle =0 
\end{equation}
and

\begin{equation}
\label{cuarenta}\langle {\cal E}_d|{\cal \vec E}_d\rangle =\langle 
{\cal \vec E}_d|{\cal E}_d\rangle =1 
\end{equation}

At degeneracy, that is, on the diabolical circle, ${\cal H}_d$ is equivalent
to a Jordan block of rank two

\begin{equation}
\label{cuarentaiuno}M^{-1}{\cal H}_dM=\left( 
\begin{array}{cc}
0 & \frac 12\Gamma \\ 
0 & 0 
\end{array}
\right) 
\end{equation}
where

\begin{equation}
\label{cuarentaidos}M=\left( 
\begin{array}{cc}
e^{-i\frac \phi 2} & 0 \\ 
ie^{i\frac \phi 2} & e^{i\frac \phi 2} 
\end{array}
\right) 
\end{equation}

Therefore, although ${\cal H}$ and ${\bf H}$ are continuous functions
of the parameters $\left( X,Y,Z\right) $ for all values of $X$ $,Y$ and $Z,$
both the Jordan normal forms and the similarity transformations leading to
them are discontinuous functions of $X$ $,Y$ and $Z$ on all points on the
diabolical circle.

In brief, we have shown that, in the case of a resonance degeneracy of rank
two, close to the crossing, the energy surface has two pieces which lie in
orthogonal subspaces. The surface which represents the real part of the
energy has the shape of an open sandglass or diabolo, with its waist at the
diabolical circle. The surface representing the imaginary part of the energy
is a sphere. The two surfaces are
embedded in orthogonal subspaces but touch each other at all points on the
diabolical circle. To the points on the diabolical circle correspond
degenerate matrices with one Jordan block of rank two, and to the points off
the diabolical circle correspond matrices with simple, that is,
non-degenerate eigenvalues. It is convenient to recall that when there are
multiple eigenvalues, the reduction of a matrix to the Jordan normal form is
not a stable operation. Indeed, in the presence of multiple eigenvalues an
arbitrarily small change in the matrix may change the Jordan normal form
completely. However, when we are dealing with a family of matrices depending
on parameters, multiple eigenvalues are unremovable by a small perturbation.
In this latter case, we can reduce every individual matrix of the family to
a Jordan normal form, but, both this normal form and the transformation
leading to it depend discontinuously on the parameters. Therefore, the
diabolical circle is a continuous line of singularities of the family of
matrices, unremovable by a small perturbation.

\section{Berry Phase of a Resonant State}

\subsection{Geometric Phase of a Resonant State.}

After having examined the topology of the energy surfaces close to a
crossing of resonances, let us go back to the expansion of the wave function 
$\Psi $ of the perturbed system in terms of bound and resonant energy
eigenstates of the unperturbed Hamiltonian,

\begin{equation}
\label{cuarentaitres}\Psi =\sum_m|\varphi _m(\xi _i)\rangle
a_m(t)+\int_c|\varphi ^{(+)}(k;\xi _i)\rangle b(k,t)dk, 
\end{equation}
the summation runs over all bound and resonant states of $H_{0.}$ As before,
in the following we will disregard the non-resonating background due to the
integral over the continuum of scattering wave functions of complex wave
number. In this approximation

\begin{equation}
\label{cuarentaicuatro}H\Psi =\sum_m\{|\varphi _m(\xi _i)\rangle \sum_n{\bf H} 
_{m,n}\left( C_i(t)\right) a_n(t)\}. 
\end{equation}
When the collective parameters change slowly with time, the perturbation
term $H_1\left( C_i(t)\right) $ is an implicit function of time. In this
case the complex non-Hermitian matrix ${\bf H}$ is also an implicit function
of time

\begin{equation}
\label{cuarentaicinco}{\bf H}_{m,n}={\cal E}_m^{(0)}\delta _{m,n}+\langle
\varphi _m|H_1(C_i(t))|\varphi _n\rangle 
\end{equation}

For values of the external parameters off the diabolical circle ${\bf H}$
has no repeated eigenvalues, in consequence, it may be brought to diagonal
form by means of a similarity transformation

\begin{equation}
\label{cuarentaiseis}{\bf K}^{-1}{\bf HK}={\bf E}, 
\end{equation}
where ${\bf E}$ is the diagonal matrix of the energy eigenvalues. The
columns in the matrix ${\bf K}$ are the instantaneous right eigenvectors of 
${\bf H}$ given by

\begin{equation}
\label{cuarentaisiete}{\bf H}\left( C_i(t)\right) |\phi ^{(s)}(t)\rangle =
{\cal \hat E}_s(t)|\phi _\lambda ^{(s)}(t)\rangle 
\end{equation}
In an obvious notation

\begin{equation}
\label{cuarentaiocho}{\bf K}=\left( |\phi ^{(1)}\rangle ,|\phi ^{(2)}\rangle
,......|\phi ^{(s)}\rangle ....|\phi ^{(n)}\rangle \right) . 
\end{equation}

The rows in ${\bf K}^{-1}$ are the corresponding left eigenvectors of ${\bf 
H}$, properly normalized.

\begin{equation}
\label{cuarentainueve}\left\langle \phi ^{(i)}|\phi ^{(j)}\right\rangle
=\delta _{ij}. 
\end{equation}

With the help of ${\bf K}$ we obtain the adiabatic basis, $\left\{ \left|
\hat \varphi _s\left( \xi ;C_i(t)\right) \right\rangle \right\} ,$ of
instantaneous bound and resonant states of the complete Hamiltonian $H$,

\begin{equation}
\label{cincuenta}|\hat \varphi _s(\xi _i;C_i(t))\rangle
=\sum_m|\varphi _m(\xi _i)\rangle {\bf K}_{m,s}(C_i(t)) 
\end{equation}
and their adjoints

\begin{equation}
\label{cincuentaiuno}\langle \hat \varphi _s(\xi _i;C_i(t)|=\sum_n({\bf K}
^{-1}(t))_{sn}\langle \varphi _n(\xi _i)| 
\end{equation}

We may now write the expansion of $\Psi $ as an expansion in instantaneous
energy eigenfunctions $\{|\hat \varphi _s(\xi _i;C_i(t))\rangle \}$ of $H$

\begin{equation}
\label{cincuentaidos}\Psi =\sum_s|\hat \varphi _s\left( \xi
_i,C_i(t)\right) \rangle \hat a_s(t) 
\end{equation}
where

\begin{equation}
\label{cincuentaitres}\hat a_s(t)=\sum_n\left( {\bf K}^{-1}(t)\right)
_{sn}a_n(t) 
\end{equation}
Similarly, the expansion of $H\Psi $ becomes

\begin{equation}
\label{cincuentaicuatro}H\Psi =\sum_s|\hat \varphi _s\left( \xi
_i;C_i(t)\right) \rangle {\cal \hat E}_s(t)\hat a_s(t) 
\end{equation}

Substitution of (\ref{cincuentaidos}) and (\ref{cincuentaicuatro}) in the
time dependent Schr\"odinger equation(\ref{uno}) gives the set of coupled
equations

\begin{equation}
\label{cincuentaicinco}\frac{d\hat a_s(t)}{dt}+\sum_{m=1}\langle \hat
\varphi _d|\nabla _R\hat \varphi _m\rangle \cdot \frac{d\vec R}{dt}\hat
a_m(t)=-i{\cal \hat E}_s(t)\hat a_s(t) 
\end{equation}

It will be assumed that the non-adiabatic transition amplitudes are very
small

\begin{equation}
\label{cincuentaiseis}\frac 1{\left| \hat a_s\right| }\left| \langle \hat 
\varphi_s|\nabla _R\hat \varphi _m\rangle \cdot \frac{d\vec R}{dt}
\hat a_m\right|<<1,\qquad m\neq s, 
\end{equation}

Then, we can make the approximation

\begin{equation}
\label{cincuentaisiete}\frac 1{\hat a_s}\frac{d\hat a_s}{dt}\simeq -i{\cal 
\hat E}_s(t)-\langle \hat \varphi _s|\nabla _R\hat \varphi _s\rangle \cdot 
\frac{d\vec R}{dt}, 
\end{equation}
Integrating both sides, we obtain

\begin{equation}
\label{cincuentaiocho}\hat a_s(t)=e^{-\frac i\hbar \int_{t_0}^t{\cal \hat E}
_s(t^{\prime })dt^{\prime }}e^{i\gamma _s}\hat a_s(0) 
\end{equation}
where the first factor is the complex dynamical phase, whereas the second
one is the complex Berry phase given by

\begin{equation}
\label{cincuentainueve}\gamma _s=i\int_{{\bf c}}\langle \hat \varphi _s|\nabla
_R\hat \varphi _s\rangle \cdot d\vec R, 
\end{equation}
in this expression ${\bf c}$ is the path traced by the system in parameter
space when $t^{\prime }$ goes from $t_0$ to $t$.

The right hand side of (\ref{cincuentainueve}) may be written as a surface
integral with the help of Stokes theorem

\begin{equation}
\label{sesenta}\gamma _s=i\sum_{m\neq s}\int_\Sigma \int_{\partial
\Sigma ={\bf c}}\langle \hat \varphi _s|\nabla _R\hat \varphi _m\rangle
\times \langle \hat \varphi _m|\nabla _R\hat \varphi _s\rangle \cdot d\vec
\Sigma 
\end{equation}
where $\Sigma $ is a surface bounded by the curve ${\bf c.}$ This expression
may be written in a more convenient form by means of the identity

\begin{equation}
\label{sesentaiuno}\langle \hat \varphi _s|\nabla _R\hat \varphi _m\rangle
=\frac 1{{\cal \hat E}_m-{\cal \hat E}_s}\langle \hat \varphi _s|\nabla
_RH_1|\varphi _m\rangle , 
\end{equation}
then,

\begin{equation}
\label{sesentaidos}\gamma _s=i\sum_{m\neq s}\int_\Sigma \int_{\partial
\Sigma ={\bf c}}\frac{\langle \hat \varphi _s|\nabla _RH_1|\hat \varphi
_m\rangle \times \langle \hat \varphi _m|\nabla _RH_1|\hat \varphi _s\rangle
\cdot d\vec \Sigma }{\left( {\cal \hat E}_s-{\cal \hat E}_m\right) ^2}, 
\end{equation}
provided the surface $\sum $ does not cross the diabolical circle where the
denominator vanishes.

\subsection{Computation of the geometric phase.}

Explicit expressions for the Berry phase $\gamma _s$ in terms of our
parametrization of the interaction Hamiltonian may easily be obtained. We
recall that the unperturbed Hamiltonian and its bound and resonant energy
eigenfunctions are independent of time. Hence, the time dependence of the
bound and resonant instantaneous energy eigenstates of the perturbed system
is entirely contained in the matrix ${\bf K,}$ which is a function of time
through the time dependence of the external parameters. Therefore, from eq.
(\ref{cincuenta})

\begin{equation}
\label{sesentaitres}|\nabla _R\hat \varphi _s(\xi _i;C_i(t))\rangle
=\sum_m|\varphi _m(\xi _i)\rangle (\nabla _R{\bf K})_{ms}. 
\end{equation}

When this expression is substituted in (\ref{cincuentainueve}), and use is 
made of the biorthonormality of the set of unperturbed bound and resonant
eigenfunctions, we get

\begin{equation}
\label{sesentaicuatro}\gamma _s=i\int_{{\bf c}}\left[ {\bf K}^{-1}(\nabla _R
{\bf K})\right] _{ss}\cdot d\vec R. 
\end{equation}
The geometric phase is now written as a path integral of the diagonal
elements of the product of the inverse and the gradient (in parameter space)
of the matrix ${\bf K,}$ which diagonalizes the matrix ${\bf H}$. This
expression for $\gamma _s$ is equivalent to (\ref{cincuentainueve}).
In the case of a finite number of interfering resonant states which may
become degenerate, all other bound or resonant energy eigenstates being
non-degenerate, we may safely assume that ${\bf H}$ is diagonal except for a
square diagonal block which mixes the interfering resonances. In this case 
${\bf K}$ is a finite matrix and the evaluation of the integrand in the right
hand side of (\ref{sesentaicuatro}) involves only the product of two finite
matrices and no integration over the particle coordinates of the microscopic
internal components of the system is involved, as would seem to be the case
in the evaluation of $\langle \hat \varphi _s|\nabla _R\hat \varphi
_s\rangle .$

Furthermore, from (\ref{cuarentaiocho}) we see that the columns in $\nabla _R
{\bf K}$ are the gradients of the instantaneous right eigenvectors of 
${\bf H}$. Therefore, $\gamma _s$ may also be written as

\begin{equation}
\label{sesentaicinco}\gamma _s=i\int_{{\bf c}}\left\langle \phi _s|\nabla
_R\phi _s\right\rangle \cdot d\vec R 
\end{equation}

From (\ref{sesentaicuatro}) we may derive a nice sum rule for the geometric
phases of the interfering resonante states,

\begin{equation}
\label{sesentaiseis}\sum_s\gamma _s=i\int tr\left[ {\bf K}^{-1}\left( 
\nabla _R{\bf K}\right) \right] \cdot d\vec R. 
\end{equation}
This is a topological invariant, namely, the first Chern class\cite
{twentytwo}, as will be shown below in the case of two interfering states.

Let us consider now the particular case of a system with two resonant states
strongly mixed by the external interaction, which may become degenerate by a
small variation of the external parameters while all other bound or resonant
states remain non-degenerate. In this case the system is in a domain in
parameter space which contains one and only one closed line of singularites
of ${\bf K}$ which is topologically equivalent to the diabolical circle
corresponding to an accidental degeneracy of the two interfering resonant
states. Then, we may safely assume that ${\bf H}$ is diagonal except for a 
$2\times 2$ block ${\bf H}_{2\times 2}$. The matrix ${\bf K}$ which
diagonalizes ${\bf H}_{2\times 2}$ is given in (\ref{treintaiseis})

The matrices ${\bf K}^{-1}$ and $\nabla _R{\bf K}$ are readily obtained from
(\ref{catorce}), and (\ref{treintaiseis})

\begin{equation}
\label{sesentaisiete}{\bf K}^{-1}=\frac 1{\sqrt{2\epsilon }}\left( 
\begin{array}{cc}
\sqrt{\epsilon +\eta } & \sqrt{\epsilon -\eta }\frac \xi {\left| \xi \right|
} \\ \sqrt{\epsilon -\eta } & -\sqrt{\epsilon +\eta }\frac \xi {\left| \xi
\right| } 
\end{array}
\right) 
\end{equation}
and

\begin{equation}
\label{sesentaiocho} 
\begin{array}{c}
\nabla 
{\bf K}=\frac 1{\sqrt{2\epsilon }}\{\left( -\frac 1{\sqrt{2\epsilon }}{\bf K}
+\frac 1{2\epsilon }{\bf M}_1\right) \left( \vec R-i\frac 12\vec \Gamma
\right) +\frac 1{2\Gamma }{\bf M}_2\vec \Gamma \\ +i\frac 1{\Gamma \left(
\epsilon ^2-\eta ^2\right) }\frac{\xi ^{*}}{\left| \xi \right| }M_3\left(
\vec \Gamma \times \vec R\right) \} 
\end{array}
\end{equation}
where the matrices M$_1,M_2$ and $M_3$ are

\begin{equation}
\label{sesentainueve}\frac 12\left( {\bf M}_1+{\bf M}_2\right) =\frac 1{\sqrt{
\epsilon +\eta }}\left( 
\begin{array}{cc}
1 & 0 \\ 
0 & -\frac{\xi ^{*}}{\left| \xi \right| } 
\end{array}
\right) , 
\end{equation}

\begin{equation}
\label{setenta}\frac 12\left( {\bf M}_1-{\bf M}_2\right) =\frac 1{
\sqrt{\epsilon -\eta }}\left( 
\begin{array}{cc}
0 & 1 \\ 
\frac{\xi ^{*}}{\left| \xi \right| } & 0 
\end{array}
\right) 
\end{equation}
and

\begin{equation}
\label{setentaiuno}{\bf M}_3=\left( 
\begin{array}{cc}
0 & 0 \\ 
\sqrt{\epsilon -\eta } & -\sqrt{\epsilon +\eta } 
\end{array}
\right) 
\end{equation}

Then, a straightforward calculation gives

\begin{equation}
\label{setentaidos}\gamma _1=-\frac 12\int_{{\bf c}}\frac 1{\Gamma
\epsilon \left( \epsilon +\eta \right) }\left( \vec \Gamma \times \vec
R\right) \cdot d\vec R 
\end{equation}
and

\begin{equation}
\label{setentaitres}\gamma _2=-\frac 12\int_{{\bf c}}\frac 1{\Gamma \epsilon
\left( \epsilon -\eta \right) }\left( \vec \Gamma \times \vec R\right) \cdot
d\vec R 
\end{equation}

These expressions are very similar to the well known results obtained for
the geometric phase of bound states\cite{one}. An obvious difference is that
the geometric phase of resonant states is complex since $\epsilon $ and 
$\eta $ are complex functions of the parameters $\vec R$ and $\vec \Gamma $.
There is another important but less apparent difference: In the case of an
accidental degeneracy of resonances $(\Gamma \neq 0),$ the denominator in
the right hand side of (\ref{setentaidos}) and (\ref{setentaitres}) vanishes on
the continuous line of singularities we have called the diabolical circle,
and not at one isolated point as is the case for bound states $(\Gamma =0)$.
It follows that two kinds of non-trivial, topologically inequivalent closed
paths are possible, as shown in Fig. 3. First, those paths which surround or
encircle the diabolical circle but are not linked to it. Second, the closed
paths which are linked to the diabolical circle. Paths of the first kind are
clearly analogous to the non-trivial paths that go around the diabolical
point while paths of the second kind have no analogue in accidental
degeneracies of bound states.

For paths of the first kind it is always possible to find a surface $\Sigma $
which spans the closed path ${\bf c}$ and does not cross the diabolical
circle. Then, using Stokes theorem we may write the geometric phase $\gamma
_{s}$ as a surface integral. Computing directly from (\ref
{setentaidos}) and (\ref{setentaitres}), we get

\begin{equation}
\label{setentaicuatro}\gamma _s=\frac{\left( -1\right) ^s}2\int_{\Sigma
_s}\int_{\partial \Sigma _s={\bf c}}\frac{\left( \vec R-i\frac 12\vec \Gamma
\right) \cdot d\vec S}{\left[\left(\vec R-i\frac 12\vec \Gamma \right)^2\right
]^{\frac 32}}, 
\end{equation}
where $s=1,2.$ Since $\gamma _2$ changes into $\gamma _1$ when $\Sigma _1$
and $\Sigma _2$ are exchanged and the sign of $d\vec S$ is changed, the
normals for $\Sigma _2$ and $\Sigma _1$ should be oppositely
oriented. If we say that $\vec \Gamma $ points upwards, then $\Sigma _1$ is
above ${\bf c}$ and $\Sigma _2$ is below ${\bf c.}$

This is, of course, the same result as would have been obtained from the
general expression (\ref{sesentaidos}) and our parametrizacion of the
perturbation term in the Hamiltonian, no summation over intermediate states
occurs in (\ref{setentaicuatro}) since in the simple case of only two
interfering resonant states the summation in (\ref{sesentaidos}) 
has only one term.

Adding $\gamma _1$ and $\gamma _2$, the sum rule is written as a surface
integral

\begin{equation}
\label{setentaicinco}\gamma _1+\gamma _2=-\frac 12\int_\Sigma \int_{\partial
\Sigma }\frac{\left( \vec R-i\frac 12\vec \Gamma \right) \cdot d\vec S}{%
\left[\left(\vec R-i\frac 12\vec \Gamma\right)^2\right]^{\frac 32}}, 
\end{equation}
where $\Sigma $ is a closed surface with the diabolical circle completely
contained in its interior, $d\vec S$ is the surface element normal to 
$\Sigma $.

The integral in (\ref{setentaicinco}) is easily computed when $\Sigma $ is a
sphere with radius $R>\frac 12\Gamma $, the result is

\begin{equation}
\label{setentaiseis}\gamma _1+\gamma _2=-2\pi 
\end{equation}

It is now easy to show that the resonance degeneracy produces a continuous
distribution of singularities on the diabolical circle. To this end, we
convert the surface integral (\ref{setentaicinco}) to a volume integral using
Gauss theorem. Then, 

\begin{equation}
\label{setentaisiete}\gamma _1+\gamma _2=-\frac 12\int \int \int_V\left(
\nabla _R\cdot \frac{\left( \vec R-i\frac 12\vec \Gamma \right) }{\left[
\left( \vec R-i\frac 12\vec \Gamma \right) ^2\right] ^{\frac 32}}\right) dV, 
\end{equation}
where $V$ is the volume inside $\sum $ and bounded by it. The term in round
brackets under the integration sign vanishes when $\epsilon \neq 0.$
Therefore, the non-vanishing value of $\gamma _1+\gamma _2$ implies the
occurrence of $\delta $-function singularities of the integrand on those
points where $\epsilon $ vanishes$.$

Hence,

\begin{equation}
\label{setentaiocho}\nabla _R\cdot \left[ \frac{\vec R-i\frac 12\vec \Gamma 
}{\left[ \left( \vec R-i\frac 12\vec \Gamma \right) ^2\right] ^{\frac 32}}
\right] =-\frac{\delta \left( R-\frac 12\Gamma \right) }{R^2}\delta \left(
\cos \theta \right) 
\end{equation}
the factor $R^{-2}$ multiplying the delta function is needed to reproduce
the value $2\pi$ of $\gamma _1+\gamma _2$.

\vspace{0.3cm}

We may say that, instead of having the fictitious magnetic charge on the
$\it monopole$ singularity characteristic of the accidental degeneracy of
bound states, in the case of an accidental degeneracy of resonant states the
fictitious magnetic charge is evenly and continuously distributed on the
closed line of singularities we have called the diabolical circle.

Now, let us consider the difference between geometric phases of resonant and
bound states. With this purpose in mind we rewrite (\ref{setentaidos}) and
(\ref{setentaitres}) as

\begin{equation}
\label{setentainueve}\gamma _{1,2}=-\frac 12\int_c\frac{\left( \vec \Gamma
\times \vec R\right) \cdot d\vec R}{\Gamma \left( \epsilon ^2-\eta ^2\right) 
}\pm \frac 12\int_c\frac{\eta \left( \vec \Gamma \times \vec R\right) \cdot
d\vec R}{\Gamma \epsilon \left( \epsilon ^2-\eta ^2\right) } 
\end{equation}

These expressions take a simple and transparent form when we change from
cartesian coordinates $\left( X,Y,Z\right) $, with $OZ$ parallel to $\vec
\Gamma $, to spherical coordinates $\left( R,\theta ,\varphi \right) $ in
parameter space. Then, (\ref{setentainueve}) becomes

\begin{equation}
\label{ochenta}\gamma _{1,2}=-\frac 12\int_{{\bf c}}d\varphi \mp \frac
12\int_{{\bf c}}\frac{\left( R\cos \theta -i\frac 12\Gamma \right) d\varphi 
}{\sqrt{R^2-\frac 14\Gamma ^2-i\Gamma R\cos \theta }} 
\end{equation}
the path $c$ is specified when $R$ and $\theta $ are given as functions of 
$\varphi .$

The sum rule for the geometric phases takes the form

\begin{equation}
\label{ochentaiuno}\gamma _1+\gamma _2=-\int_{{\bf c}}d\varphi 
\end{equation}

This result is valid for all paths ${\bf c}$, it is also valid for bound or
resonant states.

For closed paths of the first kind which go once around the diabolical circle

\begin{equation}
\label{ochentaidos}\gamma _1+\gamma _2=-2\pi \qquad \qquad {\bf c}^{(I)} 
~ of~ first~ kind 
\end{equation}

This result is valid for all paths ${\bf c}^{(I)}$, independently of the
shape of ${\bf c}^{(I)}$. It is of course, the same result we obtained using
the surface integral representation.

Still in the case of closed paths of the first kind, we may rearrange (\ref
{ochenta}) as

\begin{equation}
\label{ochentaitres}\gamma _{1,2}=-\frac 12\int_{{\bf c}^{(I)}}\left[ 1\mp 
\cos \theta \right] d\varphi \pm \triangle \gamma 
\end{equation}
where

\begin{equation}
\label{ochentaicuatro}\triangle \gamma =i\frac 14\Gamma \int_{{\bf c}^{(I)}} 
\frac{\left( i\frac 12\Gamma -2R\cos \theta \right) +\left( R-\sqrt{
R^2-\frac 14\Gamma ^2-i\Gamma R\cos \theta }\right) }{\sqrt{R^2-\frac
14\Gamma ^2-iR\Gamma \cos \theta }\left[ R-\sqrt{R^2-\frac 14\Gamma
^2-iR\Gamma \cos \theta }\right] }d\varphi 
\end{equation}

The first term in the right hand side of (\ref{ochentaitres}) is the well 
known expression for the Berry phase of two interfering bound states 
adiabatically transported in parameter space around an accidental degeneracy. 
The second term gives the difference of the actual Berry phase of the 
resonant state and the phase of a bound state,

\begin{equation}
\label{ochentaicinco}\gamma _{1,2}^{res}=\gamma _{1,2}^{bound}\pm \triangle
\gamma (c^{(I)}),\qquad \qquad c^{(I)}{~ of~ the~ first~ kind} 
\end{equation}
as shown in (\ref{ochentaicuatro}), $\triangle \gamma $ is proportional to 
$\Gamma $ and vanishes for vanishing $\Gamma .$

Therefore, the Berry phase acquired by a resonant state when it is
transported in parameter space around the diabolical circle in a path not
linked to it is equal to the sum of the real geometric phase a bound state
would have acquired if transported around the same path, plus a complex
correction term characteristic of resonant states.

In the case of closed paths of the second kind, that is, those paths which
are linked to the diabolical circle, there is no surface $\Sigma $ which
spans the closed path $c^{(II)}$ without crossing the diabolical circle, see
Fig 3. Therefore we may not use Stokes theorem to convert the path
integral into a surface integral. However, we may still compute the
geometric phase from the path integral,

For closed paths which are linked to the diabolical circle, the sum rule
gives zero

\begin{equation}
\label{ochentaiseis}\gamma _1+\gamma _2=-\int_{c^{(II)}}d\varphi =0,\qquad
\qquad c^{(II)}{~ of~ second~ kind} 
\end{equation}
since, in this case, the angle $\varphi $ starts out at some initial value 
$\varphi _0$ and, as the system traces the path $c^{(II)}$, it oscillates
between a minimum and maximum values and finally ends at the same initial
value $\varphi _0.$

The difference of $\gamma _1$ and $\gamma _2$ may be obtained from (\ref
{ochenta}),

\begin{equation}
\label{ochentaisiete}\gamma _{1,2}=\pm \triangle \gamma (c^{(II)}) 
\end{equation}
where $\triangle \gamma (c^{(II)})$ is given by an expression similar to (
\ref{ochentaicuatro}). There is no analogue to this case in bound states.

\section{Validity of the Adiabatic Approximation}

The present calculation of the Berry phase of a resonant energy eigenstate
is based upon the adiabatic approximation. In general, there seems to be
some incompatibility between the decay of the system, that is, the vanishing
of the signal and adiabaticity, i.e. slow motion. A rough estimation of the
validity of the adiabatic approximation in this case may be obtained from
the criterion given in A. Messiah\cite{twenty},

\begin{equation}
\label{ochentaiocho}\left| \frac{\max imum\ angular\ velocity\ of\ |\hat
\varphi _s\rangle }{\min imum\ Bohr\ frecuency\ of\ |\hat \varphi _s\rangle }
\right| \ll 1 
\end{equation}
which, in our notation means

\begin{equation}
\label{ochentainueve}\frac{\hbar \max \left\{ \left| \sum_{m\neq s}\langle
\hat \varphi _s|\frac{d\hat \varphi _m}{dt}\rangle \right| \right\} }{\min
\left\{ \left| {\cal E}_s-{\cal E}_m\right| \right\} }\ll 1 
\end{equation}

Now, recalling the relation

\begin{equation}
\label{noventa}\left\langle \hat \varphi _s|\frac{d\hat \varphi _m}{dt}
\right\rangle =\frac 1{\left( {\cal \vec E}_m-{\cal \vec E}_s\right)
}\left\langle \hat \varphi _s\left| \frac{dH_1}{dt}\right| \hat \varphi
_m\right\rangle 
\end{equation}

and calling

\begin{equation}
\label{noventaiuno}\triangle E=\min \left\{ \left| {\cal \vec E}_s-{\cal 
\vec E}_m\right| \right\} 
\end{equation}

the condition (\ref{ochentainueve}) becomes

\begin{equation}
\label{noventaidos}\frac{\hbar \max \left\{ \left| \left\langle \hat
\varphi _s\left| \frac{dH_1}{dt}\right| \hat \varphi _m\right\rangle \right|
\right\} }{\left( \triangle E\right) ^2}\ll 1 
\end{equation}

If T is some typical time of the driving Hamiltonian, such that

\begin{equation}
\label{noventaitres}\left| \left\langle \hat \varphi _s\left| \frac{dH_1}{dt}
\right|\hat \varphi _m\right\rangle \right| \simeq \frac 1T\left| \left
\langle \hat\varphi _s\left| H_1\right| \hat \varphi _m\right\rangle \right| 
\end{equation}

T should not be so long that the signal cannot be meassured. Hence, it seems
reasonable to estimate T from the largest half width of the interfering
resonances

\begin{equation}
\label{noventaicuatro}T\simeq \frac \hbar {\Gamma _{\max }} 
\end{equation}

Hence, a rough criterion for the validity of the adiabatic approximation for
unstable states would be

\begin{equation}
\label{noventaicinco}\frac{\max \left\{ \left| \left\langle \hat \varphi
_s\left| H_1\right| \hat \varphi _m\right\rangle \right| \right\} \Gamma
_{\max }}{\left( \triangle E\right) ^2}\ll 1, 
\end{equation}

where $max\left\{ \left| \left\langle \hat \varphi _s\left| H_1\right| \hat
\varphi _m\right\rangle \right| \right\}$ indicates the maximum value along
the path traced by the system in parameter space.

\section{Results and conclusions}

The purpose of the foregoing has been to discuss the adiabatic evolution of
an open quantum system in a state which is a superposition of resonant
states and evolves irreversibly due to the spontaneous decay of the unstable
states. More specifically, we studied the geometric phase acquired by the
resonant states when they are adiabatically transported in parameter space
by the mixing interaction around a degeneracy of resonances .

In the case of two resonant states mixed by a Hermitian interaction we find
two kinds of accidental degeneracies which may be characterized by the
number and lenght of the cycles of instantaneous energy eigenfunctions at
the degeneracy. In the first case there are two linearly independent
eigenfunctions belonging to the same repeated energy eigenvalue, that is,
two cycles of lenght one. In the second case there is only one degenerate
resonant eigenstate and one generalized resonant eigenstate belonging to the
same degenerate (repeated) energy eigenvalue, $i.e$. one cycle of lenght two.

Accidental degeneracies of the first kind, or first rank, give rise to one
` monopole' point singularity at a diabolical point in parameter space, as
in the case of degeneracies of negative energy bound states. In the present
case the degenerate states are bound states of positive energy embedded in
the continuum\cite{twentyone}. In degeneracies of the second kind, or second
rank, the fictitious magnetic charge is evenly and continuously distributed
on a closed line of singularities in parameter space, which is topologically
equivalent to a $\it diabolical$ circle. Only second rank degeneracies of two
resonances produce a true degenerate resonant state.

Close to a degeneracy of first rank, the hypersurfaces which represent the
real and imaginary parts of the resonance energies in parameter space are
two double cones lying in orthogonal subspaces, with their vertices located
at the same point, which for this reason might be called a double diabolical
point.

When the degeneracy is of the second rank, the topology of the energy
surfaces is different from that at a crossing of bound states. The energy
surfaces of the two resonant states that become degenerate are connected at
all points in a circle. Close to the crossing, the energy hypersurface has
two pieces lying in orthogonal subspaces in parameter space. The surface
representing the real part of the energy has the shape of a hyperbolic cone
of circular cross section, or an open sandglass, with its waist at the
diabolical circle. The surface of the imaginary part of the energy is a
sphere with the equator at the diabolical circle. The two surfaces touch
each other at all points on the diabolical circle.

In the case of two interfering resonant states, the geometric phase acquired
by the resonant states when transported around the diabolical circle in a
closed path which is not linked to it, may be written as the sum of two terms

\begin{equation}
\label{noventaiseis}\gamma _{1,2}^{res}\left( {\bf C}^I\right) =\gamma
_{1,2}^{bound}\left( {\bf C}^I\right) \pm \Delta \gamma \left( {\bf C}
^I\right) 
\end{equation}

The first term, $\gamma _{1,2}^{bound}\left( {\bf C}^I\right) $, is the real
geometric phase which a negative energy eigenstate would have acquired when
transported around a diabolical point in a closed path in the same parameter
space. The second term is complex, it gives rise to a change of the phase
and a dilation of the resonant state eigenfunction. Its imaginary part may
be positive or negative, in consequence, it may produce an amplification or
a damping of the wave function which may compensate or reinforce the
attenuation due to the imaginary part of the dynamical phase factor. For
long lived, narrow resonances, we may expect $\Delta \gamma $ to be small
compared with $\gamma ^{bound}$, since it is proportional to the ratio 
$\frac \Gamma R$ which is roughly proportional to the ratio of the width to
the real part of the resonant energies.

When the resonant states are transported in a closed path ${\bf C}^{II}$
which does not go around the diabolical circle but is linked to it, the
geometric phase they acquire is

\begin{equation}
\label{noventaisiete}\gamma _{1,2}^{res}\left( {\bf C}^{II}\right) =\pm
\Delta \gamma \left( {\bf C}^{II}\right) 
\end{equation}

Since it is not possible to find a continuous surface $\sum $ which spans
the closed path ${\bf C}^{II}$ without crossing the diabolical circle, we
can not make use of the theorem of Stokes to convert the path integral into
a surface integral. However, it may readily be computed as a path integral
from the expression

\begin{equation}
\label{noventaiocho}\Delta \gamma \left( {\bf C}^{II}\right) =\int_{{\bf C}
^{II}}\frac{\left( Z-i\frac 12\Gamma \right) \left( \vec \Gamma \times \vec
R^{\prime }\right) \cdot d\vec R^{\prime }}{\Gamma \sqrt{\left( \vec
R-i\frac 12\vec \Gamma \right) ^2}\left( X^2+Y^2\right) } 
\end{equation}

which is obtained from (\ref{ochenta}) and (\ref{ochentaiseis}). As in
the previous case, $\Delta \gamma \left( {\bf C}^{II}\right) $ is complex
and produces changes of phase and dilations in the resonant state wave
function. This case has no analogue in bound states.

The sum of the geometric phases acquired by two interfering resonant states
which are transported around a degeneracy in a closed path in parameter
space is a topological invariant, namely the first Chern class\cite
{twentytwo}. For closed paths of the first kind its value is the ``magnetic
charge'' on the diabolical circle, and it vanishes for paths of the second
kind.

In conclusion, we have shown that the Berry phase of resonant states differs
in various ways from that of bound states. It has some interesting
mathematical properties not present in the Berry phase of bound states. From
the physical point of view it is also interesting since it has a new term
which produces dilations of the wave function and may give rise to
observable effects not present in the geometric phase factors of bound
states.

\newpage\ 

{\bf FIGURE CAPTIONS.}

\smallskip\ 

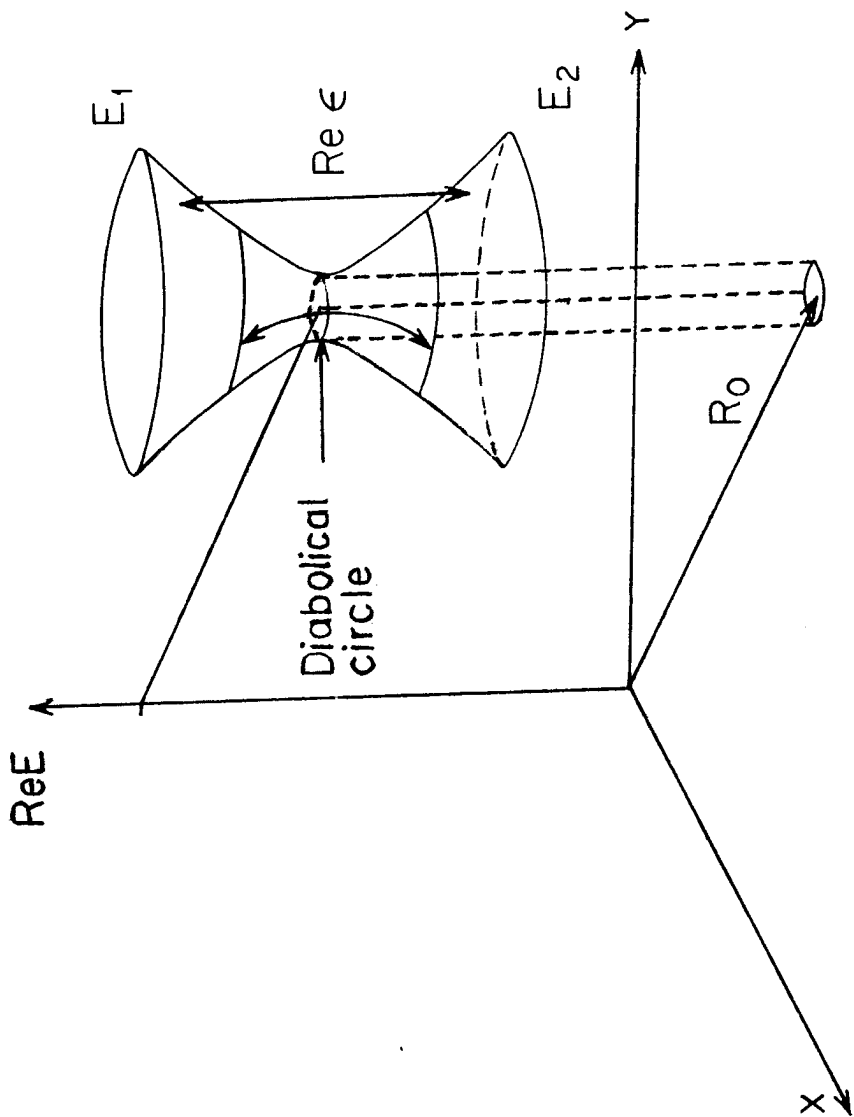
Fig. 1. Close to a crossing of resonances, the surface which represents the
real part of the energy difference in parameter space has the shape of a
hyperbolic cone of circular cross section or diabolo. The two resonances are
degenerate at the narrowest cross section or waist of the diabolo also
called the diabolical circle.

\smallskip\ 

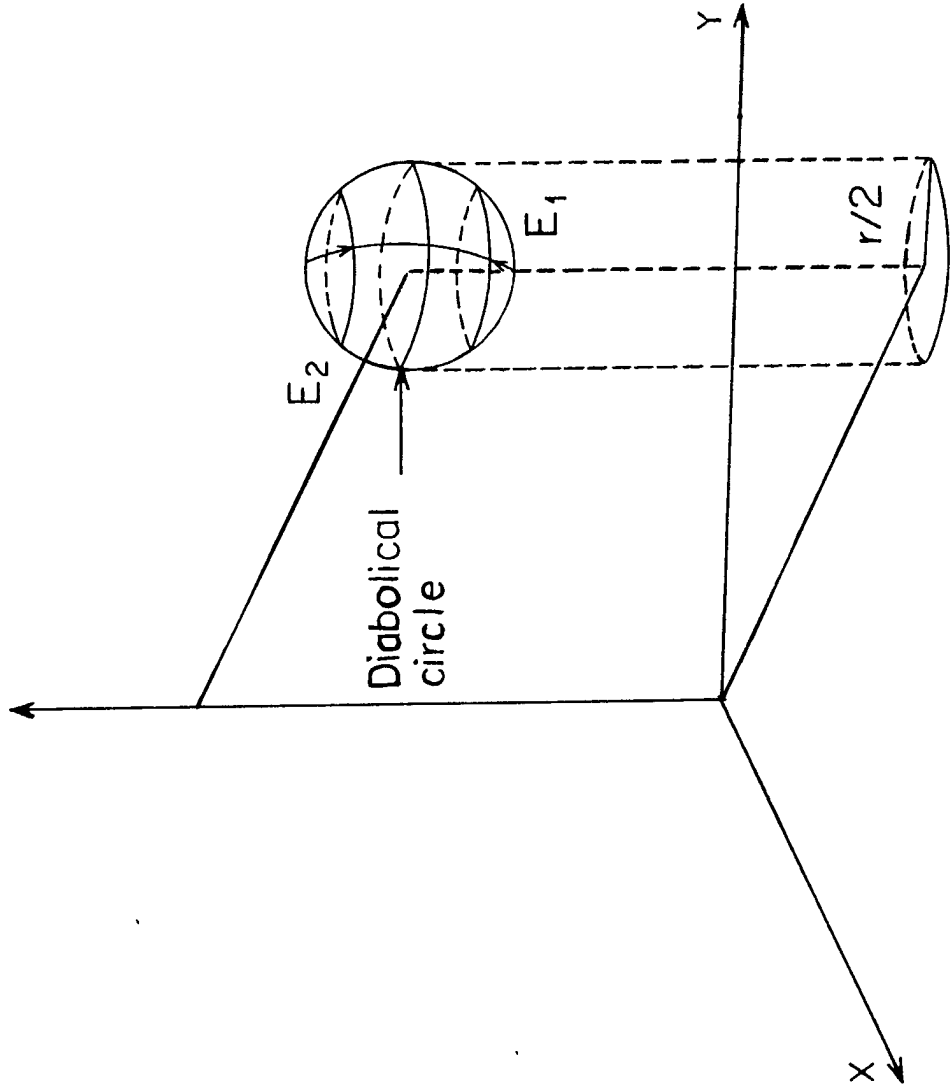
Fig. 2. The surface which represents the imaginary part of the difference of
the two energies in parameter space close to a degeneracy is a sphere. The
upper and lower hemispheres represent the imaginary parts of the two
neighbouring energies ${\bf E}_1$, ${\bf E}_2$. The equator coincides with
the diabolical circle.

\smallskip\ 

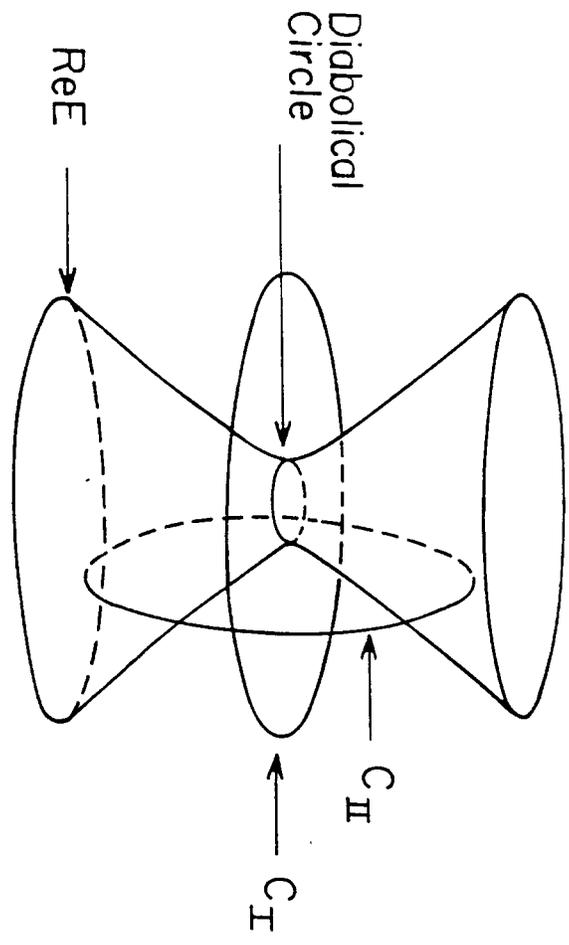
Fig. 3. In the evaluation of the Berry phase of two interfering resonances
there are two kinds of non-trivial, topologically inequivalent closed paths
in parameter space. First, those, like $C^{(I)},$ which go around the
diabolical circle but are not linked to it. Second, those, like $C^{(II)},$
which turn around the diabolical circle and are linked to it.

\end{document}